\documentclass[11pt,a4paper]{article}
\usepackage{jcappub1}
\usepackage{graphicx}
\usepackage{amssymb}
\usepackage[mathscr]{eucal}
\usepackage{amsmath}
\usepackage{afterpage}
\usepackage{rotating}
\usepackage{feynmp}
\usepackage{caption}
\usepackage{subfigure}

\title{Probing Dark Energy with Atom Interferometry.}
\author[a]{Clare Burrage,}
\author[a]{ Edmund J. Copeland}
\author[b]{and E. A. Hinds}

\affiliation[a]{School of Physics and Astronomy, University of Nottingham,
 Nottingham, NG7 2RD, UK}

\affiliation[b]{Centre for Cold Matter, Blackett Laboratory, Imperial College London, Prince Consort Road, London, SW7 2AZ, UK}

\abstract{Theories of dark energy require a screening mechanism to explain why the associated scalar fields do not mediate observable long range fifth forces. The archetype of this is the chameleon field. Here we show that individual atoms are too small to screen the chameleon field inside a large high-vacuum chamber, and therefore can detect the field with high sensitivity. We derive new limits on the chameleon parameters from existing experiments, and show that most of the remaining chameleon parameter space is readily accessible using atom interferometry.}

\notoc
\begin{document}

\maketitle
\clearpage

\section{Introduction}
The growing expansion rate of the universe, and the uneven distribution of light and matter within it, all lead to the conclusion that most of the energy in the universe is `dark energy' \cite{Copeland06}. The nature and origin of this energy are not understood.
Within quantum field theory, the natural explanation requires a new scalar field, but such a field should produce a new force \cite{Joyce:2014kja}. Laboratory and solar-system experiments show that any such `fifth force' is far weaker than gravity \cite{Adelberger:2009zz}, suggesting  in a simple Yukawa model that the underlying physics is at energies far above the Planck scale and impossible to incorporate into normal quantum field theory. This difficulty can only be avoided if the properties of the scalar field vary with environment. The archetype of this is the chameleon field \cite{Khoury:2003aq,Khoury:2003rn,Brax:2004qh}, which is screened\footnote{Screening of fifth forces may also be required in theories of modified gravity that do not attempt to explain dark energy \cite{Clifton:2011jh}.} - i.e. suppressed -  in regions of high density and so goes undetected in fifth forces experiments on earth and in the solar system \cite{Khoury:2003rn,Mota:2006fz}. 
This leaves the pressing question of how to test whether the chameleon fields actually exist.
Here we show that individual atoms, though dense in the nucleus, are too small to screen the chameleon field inside a large enough high-vacuum chamber, and therefore can detect the field with high sensitivity.  
This allows us to derive new limits on the chameleon parameters from existing experiments that measure forces on atoms. The same idea has recently been exploited by experiments probing gravitational forces with neutrons \cite{Jenke14,Brax:2011hb,Brax:2013cfa}. We go on to show that most of the remaining parameter space is readily accessible using atom interferometry to measure the chameleon force. 
Our results show that there are already more constraints on chameleon scalar fields than previously thought and open a powerful route to search for dark energy in the laboratory.

For readers unfamiliar with the chameleon  we collect together and re-derive the governing equations for the situations considered in this work in a number of appendices.  We  work in natural units where $\hbar=c=1$.  We use the $(-+++)$ metric signature.

\section{Chameleon Dark Energy}
The nature of dark energy is a central mystery in cosmology and is the thrust of major experimental activity, including the Dark Energy Survey \cite{Abbott05} and the Euclid satellite, due to be launched in 2020 \cite{Euclid12}. Chameleon theories are a significant target for these experiments. This article concerns the possibility that chameleons may be detected first in a table-top experiment on earth using ultracold atoms. Although the chameleon field $\phi$ is properly described by relativistic quantum field theory, a simple relation describes its non-relativistic steady-state: \cite{Khoury:2003aq}
\begin{equation}
\nabla^2\phi = -\frac{\Lambda^5}{\phi^2}+\frac{\rho}{M}\;,
\label{eq:effpot}
\end{equation}
where $\rho$ is the local density of matter, and we take $c=\hbar=1$. Power laws other than $1/\phi^2$ are possible, but this is a representative choice that captures all the physics \cite{{Mota:2006fz}}. In a homogeneous region, $\nabla^2\phi = 0$ and the equilibrium vacuum value of the scalar field is $\phi_{\rm eq} =(\Lambda^5M/\rho)^{1/2}$. Thus, the field is suppressed in regions of high density, and hence the force between particles - being related to the gradient of this field - is also suppressed, making it difficult to detect  near objects such as stars, planets, and laboratory test masses.

\begin{figure}[t]
\centering
\includegraphics[]{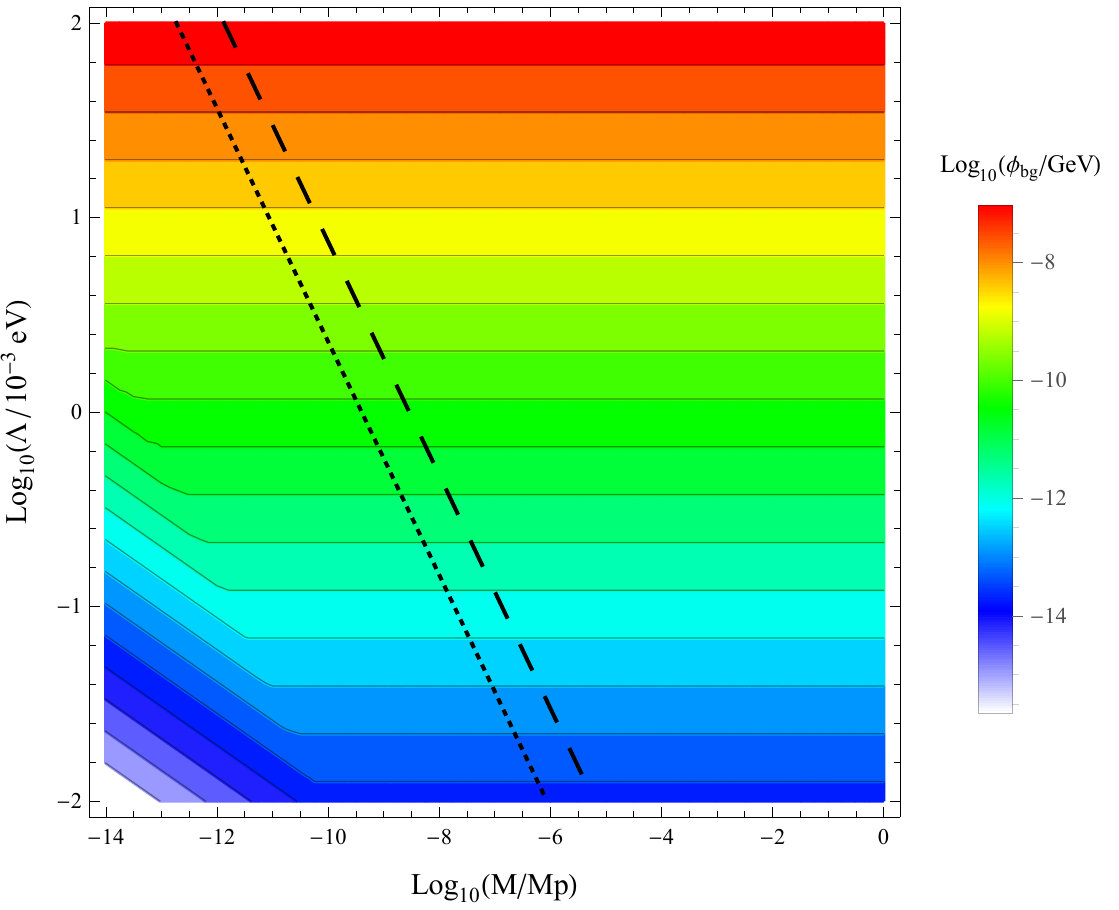}
\caption{Contour plot showing the value of $\phi_{\rm bg}$, the chameleon field at the centre of a spherical vacuum chamber, as a function of $\Lambda$ and $M$, the two parameters that characterise the field. The chamber has a radius of $10\mbox{\,cm}$ and contains $10^{-10}\mbox{\,Torr}$ of hydrogen.  In the bottom left corner $\phi_{\rm bg}$ reaches the equilibrium value $\phi_{\rm eq}=(\Lambda^5M/\rho)^{1/2}$, while above the dogleg, $\phi_{\rm bg}$ is limited by the finite size of the chamber to the lower value $0.69( \Lambda^5 L^2)^{1/3}$, which is independent of $M$. The attraction between two  bodies inside the vacuum depends on the the screening factors $\lambda$, given in Eq.(\ref{eq:lambda}). Above the dashed line, $\lambda=1$ for a caesium atom, and the force is unscreened by the atom. The dotted line is for a lithium atom. Other atoms that one might use are intermediate between these extremes.}
\label{fig:phi0}
\end{figure}

There are two coupling constants: $\Lambda$ sets the strength of the self interaction, and $M$ controls the coupling between the chameleon and matter. Through the connection to dark energy and the increasing expansion rate of the universe, $\Lambda$ is expected to be of order $1\mbox{\,meV}$ \cite{Planck14}, while Casimir force measurements indicate  $\Lambda <100 \mbox{\,meV}$ \cite{Mota:2006fz,Gannouji:2010fc,Upadhye:2012qu}. Given these constraints we take $10^{-2} \mbox{\,meV}<\Lambda<10^{+2} \mbox{\,meV}$ as our  range  of plausible values.  In comparison with this, $M$ is very poorly constrained. A lower bound of $10^4\mbox{\,GeV}$ is obtained from the measured 1s-2s interval in hydrogen \cite{Brax:2010gp}.  For the upper bound we take the reduced Planck mass $M_P\simeq2\times10^{18}\mbox{\,GeV}$, motivated by laboratory and astrophysical tests of gravity \cite{Mota:2006fz,Gannouji:2010fc,Upadhye:2012qu,Jain:2012tn}, and by the lack of clarity about physics above the Planck scale.
 In Fig.\,\ref{fig:phi0} we display this enormous area of parameter space that remains available to the chameleon. The possibility of coupling the chameleon to photons has also been explored \cite{Steffen:2010ze,Rybka:2010ah}, but this does not provide direct information about either $\Lambda$ or $M$. Other terrestrial, astrophysical and cosmological tests of gravity do not restrict the parameter space further, because of systematic uncertainties and the efficacy of the screening mechanism. 

\section{Searching for the Chameleon}
Consider $\phi$ in a typical vacuum chamber, with stainless steel walls a few mm thick, assumed spherical (for simplicity) with radius $L$. The chameleon field rises from near zero at the dense walls to a high value $\phi_{\rm bg}$ in the tenuous gas at the centre. If the chamber is large enough $\phi_{\rm bg}$ reaches the equilibrium value $\phi_{\rm eq}$, while for small chambers $\phi_{\rm bg}$ has a lower value of $0.69\times(\Lambda^5 L^2)^{1/3}$ (see Appendix \ref{sec:inside}). 
Figure\,\ref{fig:phi0} plots $\phi_{\rm bg}$ versus $\Lambda$ and $M$ for a $10\,\mbox{\,cm}$-radius chamber with  $10^{-10}\mbox{\,Torr}$ of residual hydrogen gas pressure - typical of the chambers used in cold atom experiments. In the bottom left corner of Fig.\,\ref{fig:phi0}, $\phi_{\rm bg}\rightarrow \phi_{\rm eq}$ and so depends on both $\Lambda$ and $M$, while $\phi_{\rm bg}$ elsewhere is independent of $M$, being limited by the size of the chamber.  It is clear that over a large region of the available chameleon parameter space $\phi_{\rm bg} \neq \phi_{\rm eq}$.

Now, let us place a source object $A$ and a test object $B$ near the middle of the chamber, both being small compared with the chamber. As shown in Appendix \ref{sec:chamextended}, the force between uniform spheres, due to the combined effect of gravity and the chameleon field, is \cite{Hui:2009kc} 
\begin{equation}
F_r = \frac{G M_A M_B}{r^2}\left[1 + 2 \lambda_A \lambda_B \left(\frac{M_P}{M}\right)^2\right]\;,
\label{eq:force}
\end{equation}
where $G$ is Newton's constant, $M_A$ and $M_B$ are the masses of the two objects,  $r$ is the distance between their centres of mass, and  $M_P= 1/\sqrt{8 \pi G}$ is the reduced Planck mass. The first term in equation (\ref{eq:force}) is the gravitational contribution, and the second is due to the chameleon. The coefficients $\lambda_A$ and $\lambda_B$ indicate how strongly the chameleon field is screened by each object.  These parameters are given by:
\begin{equation}
\lambda_i = \left\{ \begin{array}{lr}
1 & \;\;\;\;\;\rho_i R_i^2 < 3M \phi_{\rm bg} \;,\\
\frac{3 M \phi_{\rm bg}}{\rho_i R_i^2} & \;\;\;\;\; \rho_i R_i^2 > 3M \phi_{\rm bg}\;,
\end{array} \right.
\label{eq:lambda}
\end{equation}
where $\rho_i$ and $R_i$ are the density and radius respectively of object $i$. When $\rho_i R_i^2 > 3M \phi_{\rm bg}$, the field is suppressed inside the body, except for a thin shell near the surface, and hence the chameleon force  is reduced in comparison with the gravitational contribution in equation (\ref{eq:force}). When $\rho_i R_i^2 < 3M \phi_{\rm bg}$, the field remains essentially unsuppressed, even at  the centre of the body, and $\lambda\rightarrow1$.  We note that, when $\lambda_B =1$, the chameleon force on object $B$ takes the simple form $-\frac{M_B}{M}\vec{\nabla}\phi$, allowing us in that case to think of  $\frac{M_B}{M}\phi$ as a potential energy for the interaction.

If we suppose that $\lambda_A=\lambda_B=1$, Eq.\,(\ref{eq:force}) allows the chameleon force to be very large in comparison with the gravitational attraction because $M$ may be far below the Planck mass. However, fifth-force experiments to date have both $\lambda_A\ll1$ and $\lambda_B\ll1$,  because the objects used are large and dense, and $\phi_{\rm bg}$ is small in the high terrestrial background density. The resulting double suppression of the force is so strong that the bounds imposed by experiment are not stringent. Our central point is that one can achieve $\lambda_B=1$ using an atom in high vacuum, where $\rho_B R_B^2$ can be small, compared with $M\phi_{\rm bg}$. The acceleration towards a macroscopic test mass is then only singly suppressed, and atom interferometry is easily able to detect it.  By considering the quantity $\rho_B R_B^2$, one finds that $\lambda_B$ for the atom is determined by the nuclear density and radius, with screening by the electron cloud being insignificant in comparison.  Above the dashed line in Fig.\,\ref{fig:phi0}, $\lambda_B=1$ for a caesium atom. The dotted line is for lithium atoms.

Atoms in high vacuum have already been used to measure gravitational forces with  high precision, e.g. \cite{Fixler07,Lamporesi08}, but with source masses that are outside the vacuum chamber. Because of the intervening vacuum wall, the chameleon field within the chamber is essentially unaffected by the external  source, in close analogy with  Faraday shielding in electrostatics, as we discuss more fully in Appendix \ref{sec:screening}.  Consequently, these experiments place no useful constraints on the chameleon parameters.

\begin{figure}[t]
\centering
\includegraphics[]{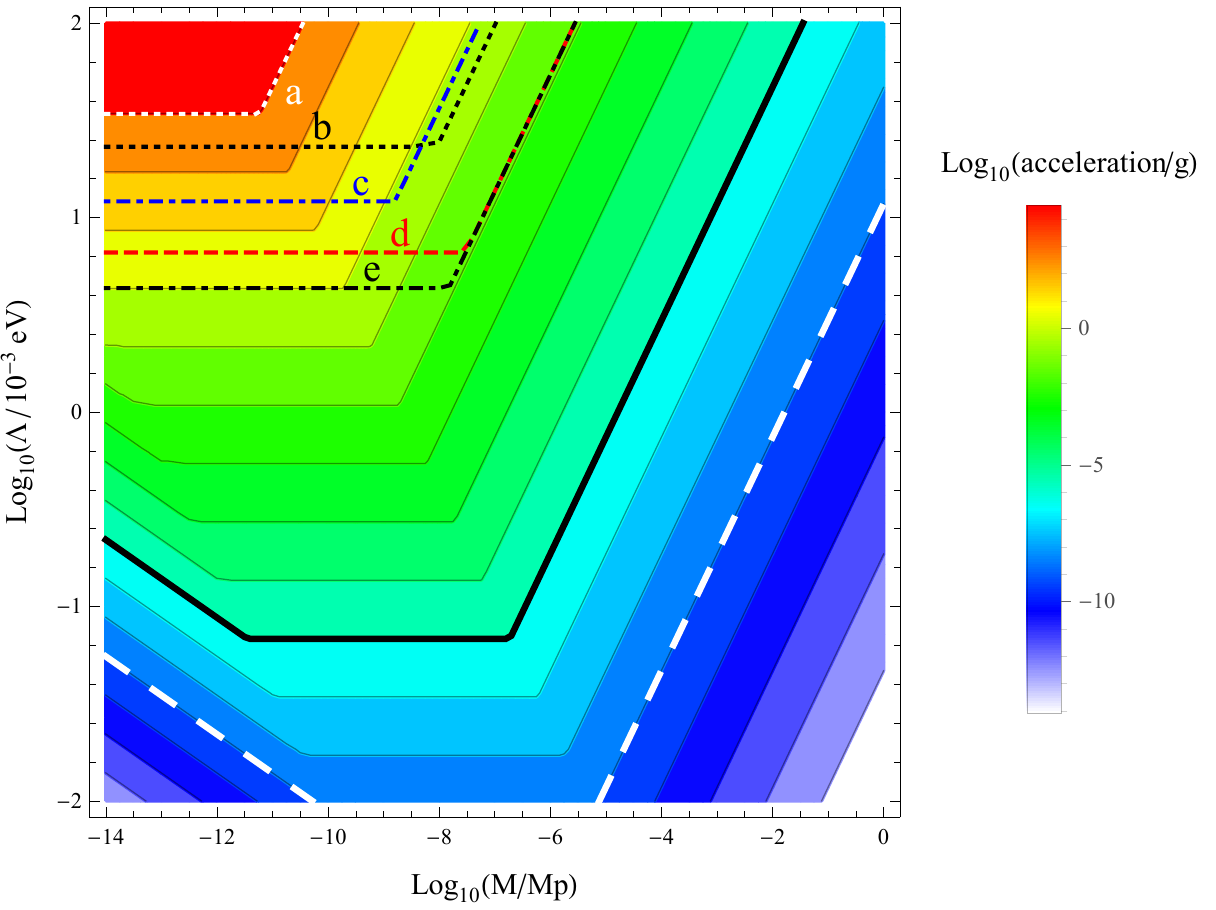}
\caption{Contour plot showing acceleration of rubidium atoms (normalised to the acceleration $g$ of free fall on earth) due to the chameleon force outside a sphere of radius $R_A=1\mbox{\,cm}$  and screening factor $\lambda_A=\frac{3 M_A \phi_{\rm bg}}{\rho_A R_A^2} $. The atom and sphere are placed centrally in a $10\mbox{\,cm}$-radius vacuum chamber containing $10^{-10}\mbox{\,Torr}$ of hydrogen (as for Fig.\,\ref{fig:phi0}). The $\Lambda-M$ area above the heavy solid black line will be excluded by a first atom interferometer experiment measuring $10^{-6}g$. With modest attention to systematic errors this can move down to the heavy dashed white line.  For $\Lambda \ge 10\,\mbox{meV}$, atom interferometry could sense chameleon physics up to the Planck mass  $M_P$. We calculate that measurements on atoms and neutrons near surfaces already exclude the top-left corner, as indicated by the light-weight lines. (a) deflection of Rb atoms by a cylinder \cite{Shih74,Shih75}. (b) Deflection of Na atoms between parallel plates \cite{Sukenik93}.  (c) Energy gradient for Rb atoms near a surface \cite{baum10}. (d) Frequency shift of harmonically trapped Rb atoms  near a plane surface \cite{Harber05}. (e) Energy shift of neutrons bouncing between plane surfaces \cite{Jenke14}. All of these contours have a sloping region at high values of $M/M_p$, where the atom/neutron is unshielded, and a flat, $M$-insensitive region where the shielding factor $\lambda_B$ falls below unity. Note the accelerations in the contour plot relate to the proposed interferometry experiment, not to the accelerations of the atoms and neutrons in experiments (a - e). }
\label{fig:exclusion}
\end{figure}

By contrast, measurements of the van der Waals force on individual alkali atoms have used macroscopic sources \emph{inside} the vacuum \cite{Shih74,Shih75,Anderson88}.  An atomic beam was fired tangentially to a 1-inch-diameter cylinder and the force was deduced from the deflection of the beam. We show in Appendix \ref{sec:chamsource} that this geometry gives a $1/r$ chameleon force, rather than the $1/r^2$ of Eq.\,(\ref{eq:force}), but otherwise the formula is very similar.  On modelling the experiment, we find an upper limit of $500\,g$ (normalised to the acceleration $g$ of free fall on earth) on the possible extra acceleration of atoms at the surface of the cylinder due a chameleon force. This excludes the $\Lambda - M$ parameter space above the dotted white line $a$ in the top left corner of Fig.\,\ref{fig:exclusion}. Ref. \cite{Sukenik93} measured the transmission of sodium atoms flying through the gap between parallel plates $0.7-7\,\mu\mbox{m}$ apart, a structure for which the scalar field has recently been calculated \cite{Ivanov13}. The measurement agrees with calculations that assume only the Casimir-Polder force, allowing us to exclude the region above line $b$. A Bose-Einstein condensate (BEC) of trapped atoms placed $130\,\mu\mbox{m}$ from an atom chip \cite{baum10} confirmed the acceleration due to gravity with a $2\,\sigma$ uncertainty of $3\,\mbox{m/s}^{2}$. Taking this as the upper limit on the chameleon force, we obtain the dot-dashed blue line $c$. We find that line $d$ marks the region excluded by measurements of  the oscillation frequency of a rubidium BEC trapped $6-9\,\mu\mbox{m}$ from a surface, which confirm the Casimir-Polder force gradient \cite{Harber05}.  Line $e$ is the boundary we calculate from the recent vibrational spectroscopy of neutrons bouncing on a surface \cite{Jenke14}. All of these contours have a sloping region at high values of $M/M_p$, where the atom/neutron is unshielded, and a flat, $M$-insensitive region where the shielding factor $\lambda_B$ falls below unity. In our analysis of the limits from the neutron experiment, we differ from Jenke \textit{et al.}\cite{Jenke14} because we take into account the weakening of the force when $\lambda_B <1$. This renders the experiment insensitive to to the chameleon fields having $\Lambda < 4\,\mbox{meV}$.  In several of these experiments, including Ref. \cite{Jenke14}, the atom or neutron is trapped in a quantum state having uncertain position. This does not alter the  shielding factor $\lambda_B$ because the size of the particle remains well defined even when the centre of mass position of the particle is uncertain.  A particle stays within a region of size $R_B$ for a time of order $R_B/v$, where $v$ is the velocity of the corresponding classical trajectory, this will typically be $v\sim 1 \mbox{ cm s}^{-1}$.  For comparison the chameleon field adapts to the arrival of a particle on the shorter timescale $\tau \sim 1/ m_{\rm min}(\rho)$, where $m_{\rm min}$ is the mass of the fluctuations about the minimum of the potential and is given by Equation (\ref{eq:mmin}).  Therefore the chameleon field adapts immediately to the arrival of a particle which is then screened, or not, as if it were static.  This is discussed further in Appendix \ref{sec:delocalised}. 

\begin{figure}[thp]
\centering
\includegraphics[scale=0.5]{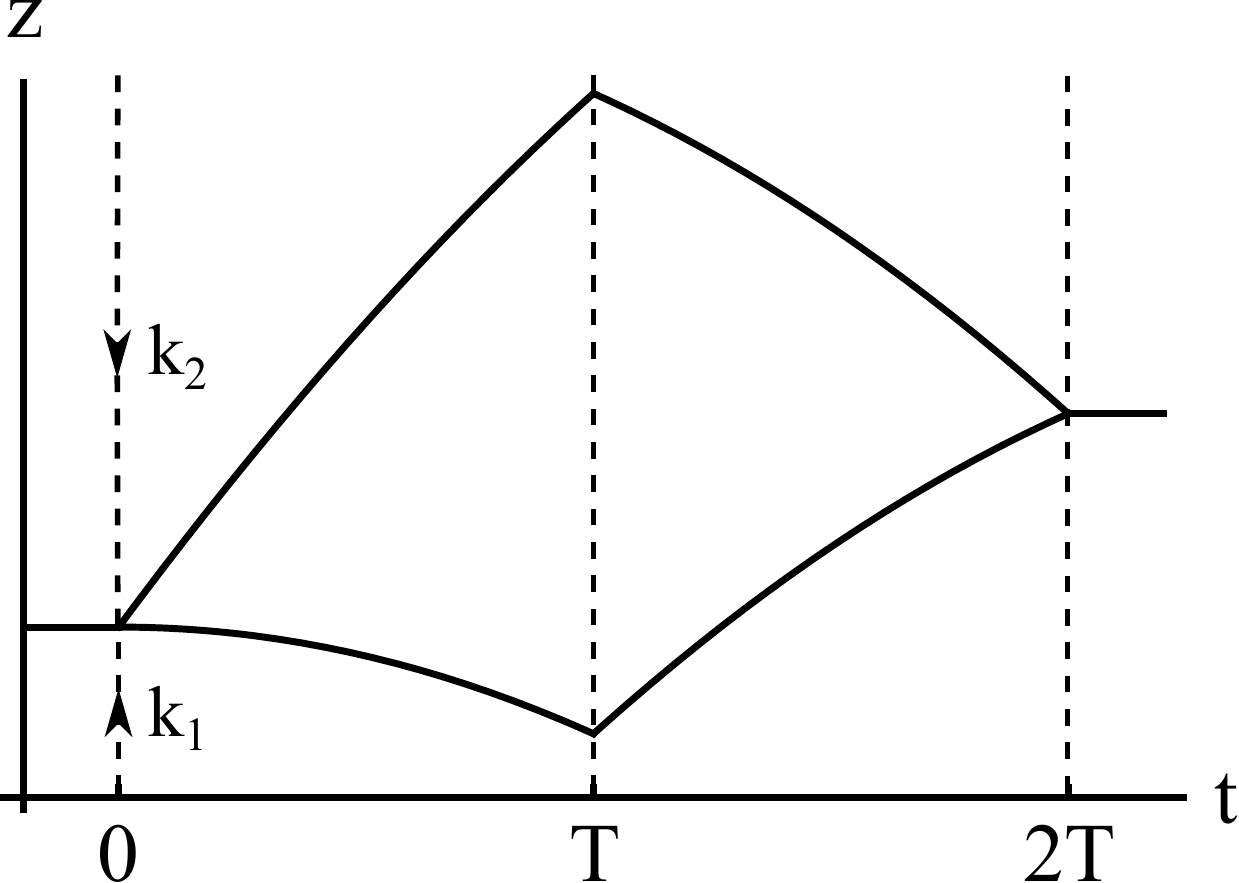}
\caption{Sketch of an atom interferometer.  Interactions between counter-propagating laser beams (dashed lines) and atoms (solid lines) can be used to give momentum to atoms and put them into a superposition of states which travel along the two arms of the interferometer.  A sequence of three pulses, separated by time $T$, is needed to split and recombine the atomic wave-function. $k_1$ and $k_2$ are the wavenumbers of the laser beams. A chameleon field gradient in the $z$ direction curves the trajectories of the atoms, and this determines  the probability of observing the atom to be in a given state at the output of the interferometer.  }
\label{fig:interferometer}
\end{figure}

\begin{figure}[htp]
\centering
\includegraphics[scale=1]{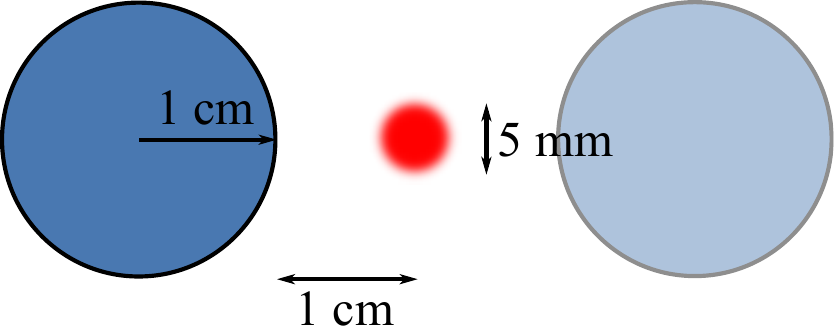}
\caption{Outline of the proposed experimental set up.  The Rubidium atoms move within the red region at the centre of the Figure. The source mass, indicated by the blue circle, is moved from its initial position on one side of the cloud of atoms, to its mirror image, indicated by the shaded blue circle. }
\label{fig:drawing}
\end{figure}

It will be much more sensitive to measure the chameleon force by interferometry of atoms in free 
fall.
  For example,  Raman interferometry \cite{Kasevich91} uses a pair of counter-propagating laser beams, pulsed on 
 three times, to split the atomic wavefunction, 
imprint a phase difference, and 
 recombine the wavefunction, as shown in Figure \ref{fig:interferometer}. 
 The output signal of the 
interferometer
 is 
 proportional 
 to $\cos^2\varphi$, with  $\varphi=(\vec{k}_1-\vec{k}_2)\cdot\vec{a}\;T^{2}$, where $\vec{k}_{1,2}$ are the wavevectors of the two laser beams, $T$ is the time interval between pulses, and $\vec{a}$ is the acceleration of the atom. We propose that rubidium atoms be cooled and launched in a small fountain, so that they stay within a $5\,\mbox{mm}$ region near a cm-sized solid mass over an interval of  $60\mbox{\,ms}$, allowing $T=30\mbox{\,ms}$ between laser pulses. A cartoon of the experimental set up is shown in Figure \ref{fig:drawing}. With the $780\,\mbox{nm}$ laser wavelength appropriate for rubidium, a $10^{-6}g$ change of acceleration along the laser beams produces a $\frac{1}{7}\,\mbox{radian}$ change in the interferometer phase $\varphi$.  A shift as large as this will be evident even in an interferometer of very modest  signal-to-noise ratio, and the constraints that such a measurement will place on the chameleon are shown by the solid black line in Figure \ref{fig:exclusion}. Our proposal is to move the source mass from one side of the atom cloud to the other and look for such a shift due to the chameleon field. We have considered a range of systematic errors that could arise when the source mass is moved. The changes due to normal gravity, the Stark effect and the Zeeman effect are all negligible at this level, as are phase shifts due to scattered light and movement of the Raman beams. These systematic errors will start to limit the sensitivity to chameleon acceleration at the level of $10^{-9}g$, as will the optical phase noise and atomic shot noise in the experiment. 

Such a measurement can explore the whole range of parameters above the heavy white dashed line in Fig.\,\ref{fig:exclusion}, and therefore gives access to new physics up to very high energy. For $\Lambda \ge 10\,\mbox{meV}$, atom interferometry should be able to detect chameleon physics right up to the Planck mass  $M_P$.  Although we have focussed here on the chameleon, we expect much the same sensitivity to any scalar field whose screening has similar phenomenology, for example, the symmetron \cite{Hinterbichler:2010es}. 

\section{Conclusions}
In summary, we have calculated the chameleon force on an atom in a vacuum chamber. We have shown that external sources are shielded by the vacuum envelope, but that a force can be produced using a cm-sized source mass inside. We find that individual atoms can sense the chameleon field without screening it and are consequently very sensitive detectors of the field. We use our results to impose new limits on the chameleon parameters, derived from existing force measurements on atoms and neutrons, and we show that most of the open chameleon parameter space is within experimental reach using current methods of atom interferometry.

\section*{Acknowledgements}
We would like to thank Justin Khoury for valuable discussions during the preparation of this work. CB is supported by a Royal Society University Research Fellowship. EAH is supported by a Royal Society Research Professorship. EJC is supported in part by the UK STFC.

\appendix

\section{The Chameleon Field Around a Source}
\label{sec:chamsource}
\subsection{The Chameleon Field Around a Spherical Source}
In this Section we review the calculation of the chameleon field profile around a static, spherically symmetric source, first derived in Reference\cite{Khoury:2003rn}.
The chameleon is a scalar field, $\phi$, whose behaviour is determined by the following action:
\begin{eqnarray}
S&=&\int d^4x\;\sqrt{-g}\left[\frac{1}{16 \pi G}R -\frac{1}{2} \nabla_{\mu} \phi \nabla^{\mu} \phi -V(\phi)\right] \nonumber \\
& &+\int d^4 x \; \mathcal{L}_{(m)} (\psi_{(m)}, \Omega^{-2}(\phi)g_{\mu\nu})\;,
\label{eq:chamaction}
\end{eqnarray}
where $g_{\mu\nu}$ is the space-time metric and $R$ the associated Ricci curvature.  $V(\phi)$ is the chameleon potential
 and $S_{(m)}=\int d^4 x \; \mathcal{L}_{(m)} (\psi_{(m)}, \Omega^{-2}(\phi)g_{\mu\nu})$ is the matter action.  Matter fields, $\psi_{(m)}$ move on geodesics of the conformally rescaled metric $\tilde{g}_{\mu\nu}=\Omega^{-2}(\phi)g_{\mu\nu}$  and the function $\Omega(\phi)$ determines the coupling between the scalar and matter fields.

The scalar equation of motion that results from the  action in equation (\ref{eq:chamaction}) is
\begin{equation}
\Box\phi =\frac{\partial V}{\partial \phi} +\frac{1}{2} \left[\frac{\partial}{\partial \phi}(\ln \Omega^2)\right] T^{(m) \alpha}_{\alpha}\;.
\label{eq:eomgeneral}
\end{equation}
Where $T^{(m)}_{\mu\nu}=(2/\sqrt{-g})(\delta S_{(m)}/\delta g^{\mu\nu})$ is the energy momentum tensor of the matter fields.  For the situations considered in this article it is sufficient to approximate matter distributions as perfect fluids with  density $\rho$ and pressure $p$.  
For a static, spherically symmetric configuration sourced by non-relativistic matter the equation of motion (\ref{eq:eomgeneral}) becomes:
\begin{equation}
\frac{1}{r^2}\frac{d}{dr}\left[ r^2 \frac{d \phi(r)}{dr} \right] =\frac{d V}{d \phi} +\frac{ \rho(r)}{M}\equiv \frac{d}{d\phi}V_{\rm eff}(\phi)\;,
\label{eq:eomspsym}
\end{equation}
where 
\begin{equation}
\frac{\partial \ln \Omega^2}{\partial \phi}=- \frac{2}{M}\;,
\end{equation}
where we have assumed that the energy scale $M$ is constant. In all cases considered in this article, the value of the field will be such that $\phi / M \ll 1$. Therefore we are able to Taylor expand the coupling function $\Omega$ around $\phi = 0$ and only keep the first term in the series that is relevant in the equation of motion leading to Equation (\ref{eq:eomspsym}).
Equation (\ref{eq:eomspsym})  can be interpreted as the chameleon moving in a density-dependent potential:
\begin{equation}
V_{\rm eff}(\phi) =V(\phi) +\left(1+ \frac{\phi }{M}\right)\rho\;.
\label{eq:Veff}
\end{equation}
We specialise to a common choice of the bare chameleon potential,
 $V(\phi)=\Lambda^5/\phi$. The minimum of the corresponding effective potential, and the mass of fluctuations around this minimum are therefore:
\begin{eqnarray}
\phi_{\rm min}(\rho)&=&\left(\frac{\Lambda^5 M}{\rho}\right)^{1/2}\;, \label{eq:phimin}\\
m_{\rm min}(\rho) &=& \sqrt{2}\left(\frac{\rho^3}{\Lambda^5 M^3}\right)^{1/4}\;.\label{eq:mmin}
\end{eqnarray}

The sources for the chameleon field that we study in this work are spherically symmetric and of constant density, therefore in the chameleon equation of motion the source term is 
\begin{equation}
\rho(r)=\rho_A \Theta(R_A-r)+\rho_{\rm bg}\Theta(r-R_A)\;,
\label{eq:sphereom}
\end{equation}
where $\rho_A$ and $R_A$ are respectively the  density and radius of the source  (which therefore has mass $M_A = (4/3)\pi \rho_A R_A^3$).  In addition, $\Theta(x)$ is the Heaviside step function, and $\rho_{\rm bg}$ is the density of the background environment surrounding the ball.  We assume that this environment has constant density and extends to infinity.

We now solve the equation of motion for the chameleon in a piecewise manner, by making appropriate approximations to the chameleon effective potential.
Far away from the source the scalar field will be close to its background value $\phi_{\rm bg}$. The contribution of the effective potential to the equation of motion is then well approximated by the mass term arising from a harmonic expansion of the potential $V_{\rm eff}(\phi) = V_{\rm eff}(\phi_{\rm bg}) + \frac{m_{\rm bg}^2}{2} [\phi - \phi_{\rm bg}]^2 + ...$
\begin{equation}
\frac{1}{r^2}\frac{\partial}{\partial r}\left(r^2 \frac{\partial \phi}{\partial r}\right)=m_{\rm bg}^2(\rho_{\rm bg})[\phi-\phi_{\rm bg}]\;,
\end{equation}
where $m_{\rm bg}^2 = d^2 V_{\rm eff}/d\phi^2|_{\phi_{\rm bg}}$.  Solutions to the equation of motion are:
\begin{equation}
\phi =\phi_{\rm bg} +\frac{\alpha}{r}e^{-m_{\rm bg}r} +\frac{\beta}{r}e^{m_{\rm bg}r}\;,
\label{eq:out}
\end{equation}
and the field profile must decay at infinity, implying $\beta=0$.

Inside the source ball, where the high density $\rho_A$ moves the minimum of the effective potential to a lower field value $\phi_A \equiv \phi_{\rm min}(\rho_A)$, there are two possible types of solution:
In the first case, the field $\phi$ decreases inside the ball, but remains everywhere greater than $\phi_A$.  In this regime, which we call weakly perturbing, the effective chameleon potential equation (\ref{eq:Veff}), is well approximated within the ball by $V_{\rm eff}\approx (\phi/M)\rho_A$ and we can solve the equation of motion, equation (\ref{eq:eomspsym}), to find
\begin{equation}
\phi=\frac{1}{8\pi R_A}\frac{M_A}{M }\frac{r^2}{R_A^2} +\frac{C}{r}+D\;.
\label{eq:indendom}
\end{equation}
For this solution to be valid everywhere inside the ball  it must be regular at the origin so we set $C=0$.  Matching $\phi$ and $\phi^{\prime}$ at $r=R_A$ between equations (\ref{eq:out}) and (\ref{eq:indendom}) gives
\begin{equation}
\phi=\phi_{\rm bg}-\frac{1}{8 \pi R_A}\frac{M_A}{ M}\left\{\begin{array}{lc}
3-\frac{r^2}{R_A^2}\;, & r<R_A\;,\\
2\frac{R_A}{r}e^{-m_{\rm bg}r}\;, & r>R_A\;.
\end{array}\right.
\end{equation}
Both of these expressions have been simplified by taking $m_{\rm bg}R_A\ll 1$.  For the experiments we are considering here, $R_A \sim 1\mbox{ cm}$ and $\rho_{\rm bg}$ corresponds to a good vacuum, making this approximation valid over almost all the relevant values of the parameters $\Lambda$ and $M$.  The weak perturbation is valid in the domain:
\begin{equation}
\frac{1}{4 \pi R_A}\frac{M_A}{M}\ll\phi_{\rm bg}\;.
\end{equation}

In the second type of solution, which we call strongly perturbing, the field inside the ball reaches $\phi_A$.  If this happens anywhere, it will happen near the centre, let us say within a radius $S$.  There, we can treat the effective potential, equation (\ref{eq:Veff}), as harmonic:  $V_{\rm eff} \approx V_{\rm eff}(\phi_A) +(1/2)m_A^2(\phi-\phi_A)^2$, where $m_A\equiv m_{\rm min}(\rho_A)$.  Then the solution to equation (\ref{eq:eomspsym}) within $r<S$ is
\begin{equation}
\phi=\phi_A +\frac{E}{r}\sinh m_Ar +\frac{F}{r}\cosh m_A r\;.
\end{equation}
We want the solution to be regular at the origin and thus require $F=0$.  The leading anharmonic correction to this potential is $-(1/2) m_A^2(\phi-\phi_A)^3/\phi_A$, so the harmonic approximation is valid as long as $\phi-\phi_A \ll \phi_A$.  Thus $S$ is a radius such that $\phi(S)=\phi_A(1+\epsilon)$, where $\epsilon <1$ is a suitably chosen constant.  In the region $S<r<R_A$, we can approximate the effective potential by the density term alone and therefore the field will have the form given by equation (\ref{eq:indendom}).

Now we sew all these pieces together,   matching $\phi$ and $\phi^{\prime}$ at $r=S$ to find the constants $C$ and $D$, then matching $\phi$ and $\phi^{\prime}$ again at $r=R_A$ determines the constant $\alpha$ and the radius $S$.  The result is
\begin{equation}
\phi=\left\{\begin{array}{lc}
\phi_A\;,  & r<S\;,\\
\phi_A +\frac{1}{8\pi R_A}\frac{M_A}{ M }\frac{r^3-3S^2 r +2S^3}{rR_A^2}\;, & S < r<R_A \;,\\
\phi_{\rm bg}-\frac{1}{4 \pi R_A}\frac{M_A}{M}\left(1-\left(\frac{S}{R_A}\right)^3\right)\frac{R_A}{r}e^{-m_{\rm bg}r}\;, & R_A<r\;,
\end{array}\right.
\label{eq:thinshellprofile}
\end{equation}
and
\begin{equation}
S=R_A\sqrt{1-\frac{8\pi}{3}\frac{M}{M_A}R_A \phi_{\rm bg}}\;.
\end{equation}
We have made two approximations here:  The first is $m_{\rm bg}R_A \ll1$, the same approximation that we made in the case of the weakly perturbing ball.  The second is $\phi_{\rm bg} \gg \phi_A$, which is well justified here because we are considering a ball of solid material surrounded by a vacuum.  The scalar field has the strongly perturbed profile provided $0\leq S\leq R_A$, which is equivalent to
\begin{equation}
\frac{3}{8\pi R_A}\frac{M_A}{M}\geq \phi_{\rm bg}\;.
\end{equation}
Khoury and Weltman\cite{Khoury:2003rn} called this the thin-shell regime  because the value of the scalar field drops from $\phi_{\rm bg}$ to $\phi_A$ over a thin region near the surface of the ball.

We find it convenient to write the scalar field outside the ball in a universal form for both weakly and strongly perturbing objects:
\begin{equation}
\phi=\phi_{\rm bg}- \lambda_A \frac{1}{4 \pi R_A}\frac{M_A}{M} \frac{R_A}{r} e^{-m_{\rm bg}r}
\label{eq:universal}
\end{equation}
where
\begin{equation}
\lambda_{A} = \left\{ \begin{array}{lc}
1\;, & \rho_A R_A^2< 3 M \phi_{\rm bg}\;,\\
1-\frac{S^3}{R_A^3}\approx \frac{3M\phi_{\rm bg}}{\rho_A R_A^2}\;, & \rho_A R_A^2> 3 M \phi_{\rm bg}\;.
\end{array}\right.
\label{eq:lambdaA}
\end{equation}
The parameter $\lambda_A$ determines how responsive the chameleon field is to the object.

The chameleon field pulls a point test particle towards the spherical test mass with acceleration
\begin{equation}
a_{\phi}=\frac{1}{M}\partial_{r}\phi\;.
\label{eq:accn}
\end{equation}
This may be compared with the usual (Newtonian) gravitational acceleration, $a_N = G M_A /r^2$. At the distances of interest here, $m_{\rm bg}r\ll1$,  the ratio is
\begin{equation}
\frac{a_{\phi}}{a_N}=\frac{\partial_{r}\phi}{M}\frac{r^2}{GM_A}=2 \lambda_A \left(\frac{M_P}{M}\right)^2\;,
\end{equation}
 $M_P$ is the reduced Planck mass: $M_P^2 =1/(8 \pi G)$. Since $ \left(\frac{M_P}{M}\right)^2$ is somewhere in the range $1 - 10^{28}$, there is every possibility that the chameleon force on a test mass can greatly exceed the Newtonian force, except in cases when $\lambda_A$ is exceedingly small.

\subsection{The Chameleon Field Around a Cylindrical Source}
We extend the discussion above to the case of a cylindrical source of density $\rho_{\rm cyl}$, radius $R_{\rm cyl}$ and inifinite extent in the $z$ direction.  The equation of motion for the chameleon is 
\begin{equation}
\frac{d^2\phi}{dr^2}+\frac{1}{r}\frac{d\phi}{dr}= \frac{d V}{d\phi} + \frac{\rho(r)}{M}\;,
\label{eq:cyl}
\end{equation}
where $r$ is now the radial position in cylindrical coordinates.  The density profile is $\rho(r)=\rho_{\rm cyl} \Theta(R_{\rm cyl}-r) + \rho_{\rm bg}\Theta(r-R_{\rm cyl})$.

The solution to this equation is found by following the same steps that were taken to find the chameleon field around a spherical source. Far away from the cylinder the right hand side of equation (\ref{eq:cyl})  is well approximated by $m_{\rm bg}^2 (\phi-\phi_{\rm bg})$, where  $m_{\rm bg}$ and $\phi_{\rm bg}$ have been defined previously.  This has solutions
\begin{equation}
\phi(r) = \phi_{\rm bg} + \alpha K_0(m_{\rm bg}r)+ \beta I_0(m_{\rm bg}r)\;,
\label{eq:outcyl}
\end{equation}
where $\alpha$ and $\beta$ are constants of integration and $I_{0}(x)$ and $K_{0}(x)$ are modified Bessel functions.  To ensure that  the field profile  decays as $r$ tends to infinity we set $\beta=0$.  

In the weakly perturbing case, the right hand side of equation (\ref{eq:cyl}) is well approximated inside the cylinder by $\rho_{\rm cyl}/M$.  The equation of motion then has the solution
\begin{equation}
\phi(r) = \frac{\rho_{\rm cyl}r^2}{4M}+C\ln r +D\;,
\label{eq:weakin}
\end{equation}
where $C$ and $D$ are two more constants of integration.  For this solution to be valid everywhere inside the cylinder it must be regular at the origin and therefore we require $C=0$.
Matching $\phi$ and $\phi^{\prime}$ in  equations (\ref{eq:outcyl}) and (\ref{eq:weakin}) at the surface of the cylinder $r=R_{\rm cyl}$ we find: 

\begin{equation}
\phi=\phi_{\rm bg}-\frac{\rho_{\rm cyl} R_{\rm cyl}^2}{2M}\left\{\begin{array}{lc}
\frac{1}{2} -\gamma_E -\frac{r^2}{2 R_{\rm cyl}^2}-\ln \left( \frac{m_{\rm bg}R_{\rm cyl}}{2}\right)\;, & r<R_{\rm cyl}\;,\\
K_0(m_{\rm bg}r)\;, & r>R_{\rm cyl}\;.
\end{array}\right.
\end{equation}
where $\gamma_E$ is the Euler-Mascheroni constant.  Both of these expressions have been simplified by taking $m_{\rm bg}R_{\rm cyl}\ll 1$. The weak perturbation is valid
in the domain:
\begin{equation}
\frac{\rho_{\rm cyl} R_{\rm cyl}^2 }{2M} K_0(m_{\rm bg} R_{\rm cyl}) \ll \phi_{\rm bg}\;.
\end{equation}

For the strongly perturbing solution the field reaches $\phi_{\rm cyl}\equiv \phi_{\rm min}(\rho_{\rm cyl})$ within a radius $S< R_{\rm cyl}$.  In this region we can approximate the right hand side of equation (\ref{eq:cyl}) as $m_{\rm cyl}^2 (\phi-\phi_{\rm cyl})$, where $m_{\rm cyl}\equiv m_{\min}(\rho_{\rm cyl})$.  Then the solution within $r<S$ is 
\begin{equation}
\phi(r) = \phi_{\rm cyl} +E K_0(m_{\rm cyl}r) +F I_0(m_{\rm cyl}r)\;,
\end{equation}
where $E$ and $F$ are integration constants and we set $E=0$ to insure regularity at the origin.  Just as in the spherical case this  harmonic approximation is valid as
long as $ \phi-\phi_{\rm cyl}\ll \phi_{\rm cyl}$. Thus $S$ is a radius such that $\phi(S)=\phi_{\rm cyl}(1+\epsilon)$ where $\epsilon <1$ is a suitably
chosen constant. In the region $S < r < R_{\rm cyl}$, we can approximate the effective potential by the
density term alone and therefore the field will have the form given by equation (\ref{eq:weakin}).  

We sew all of these parts of the solution together by ensuring that $\phi$ and $\phi^{\prime}$ are continuous at $r=S$ and $r=R_{\rm cyl}$.  
\begin{equation}
\phi=\left\{\begin{array}{lc}
\phi_{\rm cyl}\;,  & r<S\;,\\
\phi_{\rm cyl} +\frac{\rho_{\rm cyl}r^2}{4 M}\left( 1 - \frac{S^2}{r^2}+2 \frac{S^2}{r^2} \ln \left(\frac{S}{r}\right)\right)\;, & S < r<R_{\rm cyl} \;,\\
\phi_{\rm bg}-\frac{\rho_{\rm cyl}R_{\rm cyl}^2}{2M}\left(1-\frac{S^2}{R_{\rm cyl}^2}\right)K_0(m_{\rm bg }r)\;, & R_{\rm cyl}<r\;,
\end{array}\right.
\label{eq:thinshellprofile2}
\end{equation}
and the surface $S$ is determined by:
\begin{equation} \label{Eq:S-surf}
\frac{4 M\phi_{\rm bg}}{\rho_{\rm cyl}R_{\rm cyl}^2}= \left(1-\frac{S^2}{R_{\rm cyl}^2}\right)\left(1-2\gamma_E -2\ln \left(\frac{m_{\rm bg}R_{\rm cyl}}{2}\right)\right) +\frac{2S^2}{R_{\rm cy;}^2}\ln\left(\frac{S}{R_{\rm cyl}}\right)
\end{equation}

Whether the cylinder is weakly or strongly perturbing, as we have just seen, the field outside it takes the form of $\phi(r)=\phi_{\rm bg}-\frac{\rho_{\rm cyl} R_{\rm cyl}^2}{2M} K_0(m_{\rm bg}r)$ and the acceleration of a test object is 
\begin{equation}
a_{\phi}=-\frac{\partial_{r}\phi(r)}{M}= - \frac{\rho_{\rm cyl} R_{\rm cyl}^2m_{\rm bg}}{2M^2} K_{1}(m_{\rm bg} r )\simeq - \frac{\rho_{\rm cyl} R_{\rm cyl}^2}{2M^2r}\,.
\end{equation}
Here, the last step makes use of our usual approximation $m_{\rm bg}r\ll1$, and shows that the cylinder produces a $1/r$ force.  In this limit we can also find simplified expressions for $S$ and $a_{\phi}$ using Eq. (\ref{Eq:S-surf}) 
\begin{eqnarray}
\frac{\rho_{\rm cyl} R_{\rm cyl}^2}{2M} &=& \frac{\phi_{\rm bg}}{\ln(m_{\rm bg}R_{\rm cyl}/2)}\;,\\
1-\frac{S^2}{R_{\rm cy}^2} &=& -\frac{ 2 M \phi_{\rm bg}}{\rho_{\rm cyl}R_{\rm cyl}^2 \ln ( m_{\rm bg}R_{\rm cyl}/2)}\;.
\end{eqnarray}

\section{The Chameleon Force Between two Extended Sources}
\label{sec:chamextended}
In this section we compute the force exerted by ball A on a test object, ball B, of mass $M_B$, radius $R_B$ and density $\rho_B$,  this  discussion  was first presented in Reference\cite{Hui:2009kc}.   We assume a hierarchy of masses and sizes, $M_B\ll M_A$ and $R_B \ll R_A$, so that we can think of ball B as moving in a background field profile sourced by ball A.
Working in Newtonian gauge, we write the perturbed metric as 
\begin{equation}
ds^2 =-(1+2\Phi)dt^2 +(1-2\Psi)\delta_{ij}dx^idx^j\;.
\end{equation}

The ball A sources a profile for the chameleon scalar field $\phi_A(\vec{x})$ and the gravitational potentials $\Phi_{A}(\vec{x})$ and $\Psi_A(\vec{x})$.  Ball B superimposes perturbations in these,  that we assume are spherically symmetric about the centre of ball B.  We centre our spatial coordinates on the position of ball B so that
\begin{eqnarray}
\Phi&=&\Phi_{A}(\vec{x}) +\Phi_{B}(r)\;,\label{eq:Phi}\\
\Psi&=&\Psi_{A}(\vec{x}) +\Psi_{B}(r)\;,\\
\phi&=&\phi_{A}(\vec{x}) +\phi_{B}(r)\;.\label{eq:phi}
\end{eqnarray}
In addition we assume that over the small volume of ball B, the fields sourced by ball A are adequately approximated using the constant and linear terms of a Taylor series:
\begin{eqnarray}
\Phi_{A}(\vec{x}) &=&\Phi_{A}(0) +x^i \partial_i \Phi_{A}|_{x=0}\;,\label{eq:PhiA}\\
\Psi_{A}(\vec{x}) &=&\Psi_{A}(0) +x^i \partial_i \Psi_{A}|_{x=0}\;,\\
\phi_{A}(\vec{x}) &=&\phi_{A}(0) +x^i \partial_i \phi_{A}|_{x=0}\;.\label{eq:phi_A}
\end{eqnarray}
This split between the fields due to ball A and the fields due to ball B need only  make sense at the surface enclosing ball B across which we will shortly compute the momentum flux.

We assume that the gravitational field profiles are sourced by the distribution of matter, with a negligible contribution from the energy density stored in the chameleon scalar field. We again make the assumption that the matter distribution is well approximated by a static, non-relativistic, perfect fluid, whose pressure is negligible compared with the density.  The gravitational potentials around ball B are then
\begin{equation}
\Psi_{B} =\Phi_{B} = - \frac{GM_{B}}{r}\;.
\label{eq:PhiB}
\end{equation}
The chameleon potential sourced by ball B takes the form of equation (\ref{eq:universal})
\begin{equation} \label{eq:phi_B}
\phi=\phi_{\rm bg}- \lambda_B \frac{1}{4 \pi R_B}\frac{M_B}{M} \frac{R_B}{r} e^{-m_{\rm bg}r}\;.
\end{equation}

The momentum of ball B is
\begin{equation}
P^{\alpha}= \int_V \tau^{0\alpha} \;d^3 x\;,
\label{eq:momentum}
\end{equation}
where $V$ is the volume of the ball, $0$ denotes a time index, and $\tau^{\mu \nu}$ is the total energy momentum tensor of matter {\it and} gravity.  This  is defined by writing the Einstein equations as

\begin{equation}
R^{(1)}_{\mu\kappa}-\frac{1}{2}\eta^{\mu\kappa}R^{(1)\lambda}_{\lambda}=-8\pi G \tau_{\mu\kappa}\;,
\label{eq:Einstein1}
\end{equation}
where $R^{(1)}_{\mu\kappa}$ is  the part of the Ricci  tensor which is first order in metric fluctuations.
\begin{equation}
R_{\alpha\beta}^{(1)} = \frac{1}{2}\left(h^{\rho}_{\;\;\beta,\alpha\rho}-h_{\beta\alpha,\;\;\rho}^{\;\;\;\;\;\rho}-h^{\rho}_{\;\;\rho,\alpha\beta}+h_{\rho\alpha,\;\;\beta}^{\;\;\;\;\;\rho}\right)\;.
\label{eq:R1}
\end{equation}
 Hence
\begin{equation}
\tau_{\mu\nu}=T^{(m)}_{\mu\nu}+T^{(\phi)}_{\mu\nu}-\frac{1}{8 \pi G}G^{(2)}_{\mu\nu}\;,
\label{eq:tau}
\end{equation}
where $T^{(m)}$ is the matter energy momentum tensor, $T^{(\phi)}$ the scalar energy momentum tensor and $G^{(2)}_{\mu\nu}$ is the part of the Einstein tensor which is not first order in the metric.

The force on ball B is equal to the rate of change of the momentum in equation (\ref{eq:momentum}).  Differentiating equation (\ref{eq:Einstein1}) gives $\partial_{\nu} \tau^{\nu}_{\mu}=0$, and so
\begin{equation}
\partial_0\tau^0_{\mu}=-\partial_i\tau^i_{\mu}\;,
\end{equation}
where Roman indices span only space-like directions.
The force on ball B is therefore 
\begin{eqnarray}
F_i=\dot{P}_i&=&-\int_V \partial_j\tau^j_i \; d^3 x\;,\\
&=& -\int_S \tau_i^j n_j \;dS\;,
\label{eq:Force}
\end{eqnarray}
where the spherical surface $S$ is just outside ball B and $n$ is the unit vector normal to that surface.
 We now consider this integral in three parts corresponding to the three terms in equation (\ref{eq:tau}). 

First,  $T^{(m)}_{ij}$ is small outside the balls, so we neglect its contribution to the integral in equation (\ref{eq:Force}). 
Next we consider the chameleon contribution to the force.  The tensor $T^{(\phi)}$ is given by
\begin{equation}
T^{(\phi)j}_i=-\nabla_i\phi\nabla^j\phi +\delta_i^j\left(\frac{1}{2}\nabla_{\mu}\phi\nabla^{\mu}\phi +V(\phi)\right)\;,
\end{equation}
and from equations (\ref{eq:phi},\ref{eq:phi_A} and \ref{eq:phi_B})
\begin{equation}
\nabla_i\phi= +\partial_i\phi_{A}  + \frac{\lambda_BM_{B}}{4 \pi M }\frac{x_i}{r^3}\;.
\end{equation}
 Hence, to first order in the charge of ball B, $(\lambda_{B} M_{B}/M)$, the chameleon contribution to the force is

\begin{eqnarray}
-\int_S T^{(\phi)j}_i n_j \; dS &=&\frac{\lambda_{B} M_{B}}{4\pi M}\int_S\frac{1}{r^3} n_j\left[x_j\partial_i\phi_{A} +x_i\partial_j\phi_{A} -\delta_i^j x^k\partial_k\phi_{A} \right]\;dS\;,\nonumber\\
&=& -\frac{\lambda_{B} M_{B}}{ M} \partial_i\phi_{A}\;,
\label{eq:chamcon}
\end{eqnarray}
where we have neglected the contribution of the potential $V(\phi)= \Lambda^5 /\phi$ to $T^{(\phi)j}_i$  (it is straightforward to check that this is a good approximation).  When ball B is a small enough test particle that $\lambda_B=1$ this becomes the result anticipated by equation (\ref{eq:accn}) for the force due to the chameleon field.

Finally we need the gravitational contribution to the force in equation (\ref{eq:Force}).  It can be shown (see, for example Reference\cite{Hui:2009kc}) that

\begin{eqnarray}
G^{(2)}_{ij}&=&-2\Phi(\delta_{ij}\nabla^2\Phi-\partial_i\partial_j\Phi) -2\Psi(\partial_{ij}\nabla^2 \Psi-\partial_i\partial_j \Psi)+\partial_i\Phi\partial_j\Phi\nonumber\\
& & -\delta_{ij} \partial_k\Phi\partial^k \Phi +3 \partial_i\Psi\partial_j\Psi -2\delta_{ij}\partial_k\Psi\partial^k\Psi -\partial_i\Phi\partial_j\Psi -\partial_i\Psi\partial_j\Phi\nonumber\\
& &+2\Psi[\delta_{ij}\nabla^2(\Phi-\Psi)+\partial_i\partial_j(\Psi-\Phi)]\;.
\end{eqnarray}

From equations (\ref{eq:Phi}), (\ref{eq:PhiA}) and (\ref{eq:PhiB}) we have
\begin{eqnarray}
\partial_k \Phi &=&\partial_k \Phi_{A} +\frac{GM_{B}}{r^2}\frac{x_k}{r}\;,\\
\partial_j\partial_k\Phi &=& \frac{GM_{B}}{r^3}\left(\delta_{jk}-3\frac{x_kx_j}{r^2}\right)\;,\\
\nabla^2 \Phi &=& 0\;,
\end{eqnarray}
and similarly for $\Psi$.  Recalling that within our approximations $\partial_k \Phi_{A}$ is constant, then to 
 first order in the Newtonian potential of ball B, $(G M_{B}/r)$, we find

\begin{eqnarray}
G^{(2)}_{ij}=\mbox{Const.}_{ij} &+&\frac{GM_{B}}{r^3}\left[2(\Phi_{A}+\Psi_{A})\delta_{ij}-6(\Phi_{A}+\Psi_{A})\frac{x_ix_j}{r^2}\right.\nonumber\\
& &\left.-6(\partial^k\Phi_{A}+\partial^k\Psi_{A})\frac{x_ix_jx_k}{r^2}+2\partial_i \Psi_{A} x_j +2 \partial_j\Psi_{A}x_i \right.\nonumber\\
& &\left.-2 \delta_{ij} \partial^k\Psi_{A} x_k\right] \;,
\end{eqnarray}
where the first term is a constant that will be irrelevant for our calculation.  Using the spherical symmetry of S, this gives the gravitational force on ball B
\begin{equation}
\frac{1}{8\pi G}\int_S G^{(2)j}_i n_j \; dS = -  M_{B} \partial_i \Phi_{A}\;.
\label{eq:Gravcon}
\end{equation}

Putting equations (\ref{eq:chamcon}) and (\ref{eq:Gravcon}) together the total force on ball $B$ becomes:
\begin{equation}
F_i =-M_{B} \left( \partial_i\Phi_{A} +\lambda_{B}\frac{\partial_i\phi_{A}}{M}\right)\;.
\end{equation}
To evaluate the gradients of  $\Phi_{A}$ and $\phi_{A}$  it is convenient to switch to coordinates centred on  spherical ball A, with ball B located at radius $r$. Then
\begin{equation}
\partial_r \Phi_{A}=  \frac{G M_{A}}{r^2}\;,
\end{equation}
\begin{equation}
\partial_r\phi_{A} = \lambda_{A} \frac{  M_{A}}{4 \pi M r^2}\;,
\end{equation}
and the total attractive force between the  balls is
\begin{equation}
F_r=\frac{GM_{A}M_{B}}{r^2}\left(1+ 2\lambda_{A}\lambda_{B}\left(\frac{M_P}{M}\right)^2\right)\;,
\end{equation}
where we have used the definition of the reduced Planck mass $8 \pi M_P^2 G=1$.  
When the balls are weakly perturbing, $\lambda_A, \lambda_B \sim 1$, the chameleon force is larger than the gravitational attraction by the potentially very large factor $2(M_P/M)^2$.  However nearly all tests of gravity to date involve macroscopic objects for which $\lambda_A, \lambda_B\ll 1$.

\section{Inside  a Vacuum Chamber}
\label{sec:inside}

Experiments to measure the force on a beam or cloud of atoms require an ultra-high vacuum chamber to protect the atoms from collisions with the gas in the atmosphere.  The chameleon force depends on $\lambda_A\lambda_B$ and hence on the background value of $\phi$ (see equation (\ref{eq:lambdaA})).  So we need to determine $\phi_{\rm bg}$ inside the vacuum chamber.  Within the wall of the chamber, the density $\rho_{\rm wall}$ is high, and the scalar field has a correspondingly low equilibrium value $\phi_{\rm wall}=\phi_{\rm min}(\rho_{\rm wall})$ given by equation (\ref{eq:phimin}).  In the vacuum, where the density is $\rho_{\rm vac} \sim 10^{-16}\rho_{\rm wall}$, the scalar field rises towards a much higher equilibrium value $\phi_{\rm eq}=\phi_{\rm min}(\rho_{\rm vac})$, but we need to determine whether it can reach this equilibrium in a chamber of limited size $L$.

The field adapts to the change of density between the walls and the vacuum over a characteristic distance of $1/m_{\rm min}(\rho_{\rm vac})$.  Thus we expect the field at the centre of the chamber to adapt to the vacuum value only if the chamber is large enough to satisfy $L\gtrsim\Lambda^{5/4} M^{3/4}\rho_{\rm vac}^{-3/4}$.  With a chamber of $0.1 \mbox{ m}$ radius ($L=5 \times 10^{14} \mbox{ GeV}^{-1}$) and a vacuum of $10^{-10}\mbox{ Torr}$ of hydrogen ($\rho_{\rm vac}=10^{-14}\mbox{ kg m}^{-3}= 5 \times 10^{-35} \mbox{ GeV}^4$) this requires 
\begin{equation}
\left(\frac{\Lambda}{10^{-3}\mbox{ eV}}\right)^{5/4}\left(\frac{M}{M_P}\right)^{3/4} \lesssim 5 \times 10^{-11}\;.
\label{eq:vaccon}
\end{equation}
Equation (\ref{eq:vaccon}) is only satisfied for particularly low values of $\Lambda$ and $M$.  Over most of the parameter space, the opposite is true and therefore the scalar field is smaller than the equilibrium value $\phi_{\rm vac}$ throughout the vacuum region.  In this case we  can  neglect the last term in equation (\ref{eq:Veff}) and approximate the equation of motion over the region of the vacuum chamber as $\nabla^2 \phi = (\partial/\partial \phi)(\Lambda^5/\phi)$. 
We expect that the value of  $\phi$ at the centre of the vacuum chamber will be such that the corresponding mass of the field will be of order $1/L$, \cite{Khoury:2003rn}, implying that 
\begin{equation}
 \phi_{\rm bg} = c (\Lambda^5 L^2)^{1/3}\;,
\label{eq:guessphi} 
\end{equation}
 where the proportionality constant $c$ is determined by numerically solving the equation of motion in the vacuum chamber. To do this we first of all note that in the interior of the walls
 of the vacuum chamber $\phi\approx \phi_{\rm wall}$.  For the example vacuum chamber described above, this remains smaller than the expression on the right hand side of equation (\ref{eq:guessphi}), for the whole of the interesting parameter space. Therefore we make the approximation that $\phi(L)$ is negligible when compared to the value of $\phi_{\rm bg}$,   it then  becomes straightforward to solve the chameleon equation of motion numerically in the interior of the vacuum chamber and determine the constant of proportionality in equation (\ref{eq:guessphi}).  We obtain
\begin{equation}
\phi_{\rm bg} = 0.69 \left( \Lambda^5 L^2\right)^{1/3}
\end{equation}
This expression is valid as long as $\phi_{\rm bg}<\phi_{\rm eq}$, requiring  $M^3 \Lambda^5 > (0.69)^6 L^4 \rho_{\rm vac}^3$. When this condition is not satisfied $\phi_{\rm bg} = \phi_{\rm eq} = (\Lambda^5 M /\rho_{\rm vac})^{1/2}$.

\section{Screening Due to the Walls of the Vacuum Chamber}
\label{sec:screening}

The derivation of the chameleon field in the interior of the vacuum chamber in the previous Section relied on the assumptions that inside the walls of the chamber $\phi \approx 0$ and $\nabla \phi \approx 0$. Perturbations sourced outside the vacuum chamber can therefore only affect what occurs inside if the perturbation can render one of these two assumptions invalid.  
To see how a perturbation in the exterior of the vacuum chamber affects these assumptions let us consider a vacuum chamber as a spherical shell of density $\rho_{\rm wall}$ and exterior radius   $R$.  When the system is  unperturbed we define a second radius $S<R$ such that $S$ is the largest radius where   $\phi(S) = 0$ and $\nabla \phi(S)=0$.  In the language of Section \ref{sec:chamsource}  if such a radius exists the vacuum chamber is a strongly perturbing object.

We now introduce a perturbation to the chameleon field in the exterior of the shell that has a constant gradient oriented along the $z$-direction; $\phi_{\rm pert}= \alpha z = \alpha r \cos \theta$ where we are working in the usual cylindrical coordinates.  This is a good approximation to the external fields in a typical laboratory.   Outside the vacuum chamber  we will assume that the chameleon is sufficiently light that its mass can be  neglected over laboratory distance scales.  Therefore in the exterior of the shell the equation of motion for the chameleon is
\begin{equation}
\nabla^2 \phi =0\;, \mbox{  when  } r>R\;,
\label{eq:eq1}
\end{equation}
In the region $S<r<R$ the equation of motion becomes
\begin{equation}
\nabla^2 \phi =\frac{\rho_{\rm wall}}{M}\;.
\label{eq:eq2}
\end{equation}
We can solve  equations (\ref{eq:eq1}) and (\ref{eq:eq2}) in terms of Legendre polynomials.  By imposing that the field and its first derivative be continuous at $r=R$, that $\phi(S) = 0$ and $\nabla \phi(S)=0$, and that the field sourced by the vacuum chamber decays as $r\rightarrow \infty$ we find that: 

\begin{equation}
\phi(r) =\alpha r\cos\theta \left(1+\frac{S^3}{2r^3}\right)+\frac{\rho_{\rm wall}}{6Mr}\left\{\begin{array}{cc}
r^3 +2S^3-3R^2 r\;, &  S<r<R\;,\\
2(S^3-R^3)\;, &  R<r\;.
\end{array}\right.
\label{eq:thinshellwithpert}
\end{equation}

In the absence of the perturbation $\alpha=0$ this reproduces the chameleon field profile around a strongly perturbing source given in equation (\ref{eq:thinshellprofile}) under the assumption that $\phi_A\approx 0$. Equation (\ref{eq:thinshellwithpert}) shows that if the surface $S$ exists in the absence of a perturbation, it continues to exist at the same position in the presence of external perturbations.  Therefore the derivation of the chameleon field in the interior of the vacuum chamber remains that discussed in Section \ref{sec:inside}, and is unaffected by the exterior perturbation.  Strongly perturbing objects screen their interior from perturbations in the exterior.

This is closely analogous to the shielding of electric fields by a shell of highly conducting material, although equation (\ref{eq:thinshellwithpert}) differs from its electrostatic analogue because the density of the shell  gives rise to a monopole in the exterior chameleon field profile that is absent when considering magnetic shielding.  The fact that the exterior perturbations only penetrate a restricted distance into the interior of the shell, is analogous to skin depth effects in electrostatic shielding.

\section{The Chameleon Field Around a Delocalised Particle}
\label{sec:delocalised}

The shielding factor $\lambda_B$ for the probe particle depends on the quantity $\rho_B R_B^2$, by exact analogy with the factor $\lambda_A$ for the source particle given in Eq.(\ref{eq:lambdaA}). For the atom of mass $M_B$, we have  $\rho_B R_B^2=\frac{3 M_B}{4 \pi R_B}$, and since the mass is virtually all in the nucleus, the relevant radius is that of the nucleus. In comparison with this, the suppression of the chameleon field due to the electron density is negligible. 

So far we have computed the chameleon field around classical sources, however the experiments we discuss here utilise atoms or neutrons, whose motion may need to be described quantum mechanically.  In this section we discuss why quantum nature of the motion does not alter the screening of the chameleon field.  

In the absence of any external forces, we can write the Hamiltonian in two parts (see for example chapter 15.4 of Merzbacher  \cite{merzbacher}). One describes the non-relativistic centre-of-mass motion $H_{\rm CM}= -(\hbar^2 /2 M_n) \nabla_X^2$, and depends on the centre of mass coordinate $X$ and the mass of the nucleus $M_n$. The other part is $H_{\rm int}(x_i)$, which depends on the coordinates $x_i$ (and spin) of the constituent particles, measured relative to $X$. This describes the internal structure of the nucleus. For the ground state  we have $H_{\rm int} u(x_i) = E_0 u(x_i)$.
The internal and external coordinates separate exactly, allowing us to write the centre of mass  eigenstates $v(X)$, which satisfy the eigenvalue equation $H_{\rm CM} v(X) = E_{\rm CM} v(X)$. 
These are momentum eigenstates
$v(X) = e^{i k X}$. The total wavefunction is $\psi = u(x_i) v(X)$.

Let us now add an external perturbation $V(X)$, which includes the gravitational and chameleonic effects of the source object and any additional trapping potential. The total Hamiltonian
becomes $H = H_{\rm in t} + H_{\rm CM} + V(X)$. In principle, we should worry about possible perturbation of the
internal state by the external potential, but this is entirely negligible in the case we are considering because the energy required to excite the nucleus from its ground state is enormous in comparison with the coupling of internal states due to the external forces.  This means that the new Hamiltonian differs only in the CM part, which now has eigenvalues $w(X)_n$ given by 
\begin{equation}
\left[ -\frac{\hbar^2}{2M_n}\nabla_X^2 +V(X) \right] w(X)_n = E_n w(X)_n
\end{equation}
To summarise, the energy of the nucleus is the sum of internal and motional energies $E_0 + E_n$.
Its wavefunction is the product of internal and external functions $\psi = u(x_i) w(X)_n$.
The separation of internal and external coordinates remain valid because, as we have just indicated, the forces exerted by the
external potential on the moving nucleus are insufficient to produce any appreciable distortion of the
shape of the nucleus.

Consider now the mean square radius $r^2$ of the mass distribution inside the nucleus in this state. We might write an operator for this as $\sum_i m_i x_i^2$, where $m_i$ and $x_i$ are respectively the masses and positions of the constituent parts of the nucleus. The total mass is $M = \sum_i m_i$, and then
\begin{equation}
r^2 = \frac{1}{M} \langle w(X) | \langle u(x_i)| \sum_i m_i x_i^2 |u(x_i)\rangle |w(X)\rangle.
\end{equation}
The operator $\sum_i m_i x_i^2$ does not act on the centre of mass coordinate, therefore  this reduces to
\begin{equation}
r^2 = \frac{1}{M}  \langle u(x_i)| \sum_i m_i x_i^2 |u(x_i)\rangle.
\end{equation}
Of course that is a non-relativistic over-simplification, but it serves to make our point.
Namely, the size of the nucleus depends on the relative positions of the constituent particles, not on the position of the centre of mass.  In exactly the same way, the centre of mass motion  has no influence on the mass $M_n$. This proves that the density of matter inside the nucleus does not care about the centre of mass motion.  Therefore even when the position of the centre of mass of the nucleus is highly uncertain the mass and radius of the nucleus are well defined.

 Still, a trapped  particle explores a volume within the trap and one might reasonably ask how that exploration affects the interaction with the chameleon field. A classically trapped particle moves along a trajectory with deterministic velocity $v$. For a particle trapped in the quantum ground state, $v$ is uncertain, with a spread given by the inverse size of the region explored by the particle, in accordance with the uncertainty principle. In either case - classical or quantum - the particle remains within any given region of size $R_B$ for a time of order $R_B/v$, and for the atoms and neutrons of interest here, $v\sim 1\,\mbox{cm/s}$.  For comparison, the chameleon field adapts to the arrival of the particle at a particular place over a time $\tau\sim1/m_{\rm min}(\rho)$ given by Eq.(\ref{eq:mmin}). With all but the very largest values of $\Lambda^5M^3$, this time is much shorter than $R_B/v$, meaning that the chameleon adapts immediately to the instantaneous position of the nucleus.  Hence, the outer part of the nucleus shields the centre from the chameleon field, in accordance with  Eq.(\ref{eq:lambdaA}), as though the particle were static. This shielding, which was neglected in  Jenke \textit{et al.}\cite{Jenke14}, is responsible for the flat bottom of the excluded regions $a-e$ in Fig.\,2 of our Article. While the dip in the chameleon field is always centred on the instantaneous position of the nucleus, the mean value of the chameleon field is the convolution of this dip with the centre-of-mass distribution of the particle in the trap.

\end{document}